\documentclass[showpacs,aps,prd,reprint,superscriptaddress,nofootinbib,longbibliography]{revtex4-2}
\usepackage[colorlinks=true, pdfstartview=FitV,
bookmarks=true, bookmarksnumbered=true, breaklinks]{hyperref}
\usepackage[dvipdfmx]{graphicx}
\usepackage{amsmath,amssymb,bm,color,longtable,mathrsfs,slashed,comment,tikz}
\usepackage{braket}
\usepackage{subfigure}

\definecolor{darkviolet}{rgb}{0.58, 0.0, 0.83}
\definecolor{electricultramarine}{rgb}{0.25, 0.0, 1.0}
\definecolor{brightpink}{rgb}{1.0, 0.0, 0.5}
\hypersetup{linkcolor=brightpink, citecolor=darkviolet, urlcolor=electricultramarine}

\definecolor{lime}{HTML}{A6CE39}
\DeclareRobustCommand{\orcidicon}{
	\hspace{-3mm}
	\begin{tikzpicture}
	\draw[lime, fill=lime] (0,0) 
	circle [radius=0.16] 
	node[white] {{\fontfamily{qag}\selectfont \tiny ID}};
	\draw[white, fill=white] (-0.0625,0.095) 
	circle [radius=0.007];
	\end{tikzpicture}
	\hspace{-3mm}
}

\foreach \x in {A, ..., Z}{\expandafter\xdef\csname orcid\x\endcsname{\noexpand\href{https://orcid.org/\csname orcidauthor\x\endcsname}
			{\noexpand\orcidicon}}
}

\begin{document}

\title{Meson gravitational D-form factors and symmetry breaking in low-energy QCD}

\author{Mamiya~Kawaguchi\orcidA{}}
\email[]{mamiya@aust.edu.cn}
\affiliation{Center for Fundamental Physics, School of Mechanics and Photoelectric Physics,
Anhui University of Science and Technology, Huainan, 232001, China}

\author{Kazuhiro~Tanaka\orcidB{}}
\email[]{kztanaka@juntendo.ac.jp}
\affiliation{Department of Physics, Juntendo University, Inzai, Chiba 270-1606, Japan}

\author{Mitsuru~Tanaka\orcidC{}}
\email[]{tanaka@hken.phys.nagoya-u.ac.jp}
\affiliation{Department of Physics, Nagoya University, Nagoya 464-8602, Japan}

\begin{abstract}
We investigate meson gravitational D-form factors in the three-flavor linear sigma model and their connection to symmetry breaking in low-energy QCD.
Within the scalar meson dominance picture, we examine the scalar meson exchange contributions and their relation to meson masses through scalar-meson couplings.
In the three-flavor symmetric limit, we first derive analytical expressions for the D-form factors of pseudoscalar and scalar mesons. In the chiral limit, the forward-limit value of the octet pseudoscalar D-form factor is fixed to $-1$, whereas the $U(1)_A$ anomaly generates additional contributions to the singlet pseudoscalar and scalar-meson D-form factors.
For realistic flavor breaking, 
the interaction introduced to reproduce the scalar-meson mass hierarchy below $1\,{\rm GeV}$ significantly affects the scalar-meson D-form factors.
We further compare the linear sigma model results with those obtained from a chiral perturbation theory Lagrangian including a dilatonic scalar field and show that their differences are governed by the canonical mass dimensions of the interaction terms contributing to meson mass generation.
These results indicate that meson D-form factors provide sensitive probes of symmetry-breaking structures in low-energy hadron physics.

\end{abstract}

\maketitle

%%%%%%%%%%%%%%%%%%%%%%%%%%%%%%%%%%%%%%%%
\section{Introduction}\label{ch:introduction}

The origin of hadron masses and their internal structure remain fundamental questions in QCD. These properties are closely related to nonperturbative aspects of QCD, such as spontaneous chiral symmetry breaking and other symmetry-breaking mechanisms. Clarifying how these mechanisms are reflected in hadron structure is therefore important for understanding the low-energy dynamics of QCD. In recent years, hadron gravitational form factors (GFFs) have attracted considerable attention as probes of hadron structure (for reviews, see Refs.~\cite{Polyakov:2018zvc,Burkert:2023wzr}).

Hadron GFFs are defined through matrix elements of the energy-momentum tensor (EMT)~\cite{Pagels:1966zza} and provide information on the mass, spin, and internal mechanical structure of hadrons. They have been studied extensively using various theoretical approaches~\cite{Hudson:2017xug,Kumano:2017lhr,Polyakov:2018exb,Lorce:2018egm,Hatta:2018sqd,Tanaka:2018wea,Anikin:2019kwi,Avelino:2019esh,Yanagihara:2019foh,Hatta:2019lxo,Freese:2019eww,Azizi:2019ytx,Mamo:2019mka,Neubelt:2019sou,Krutov:2020ewr,Alharazin:2020yjv,Varma:2020crx,Kim:2020nug,Chakrabarti:2020kdc,Yanagihara:2020tvs,Kim:2020lrs,Tong:2021ctu,Tong:2022zax,Freese:2021czn,Panteleeva:2021iip,Hatta:2021can,Mamo:2021krl,Freese:2021qtb,Gegelia:2021wnj,Kim:2021jjf,Owa:2021hnj,Lorce:2021xku,Ji:2021mfb,More:2021stk,Mamo:2022eui,Lorce:2022cle,Fujita:2022jus,Choudhary:2022den,Kim:2022wkc,Alharazin:2022wjj,Won:2022cyy,Tanaka:2022wzy,Ito:2023oby,Lorce:2023zzg,Amor-Quiroz:2023rke,Guo:2023pqw,Won:2023ial,Guo:2023qgu,Czarnecki:2023yqd,Won:2023zmf,Hatta:2023fqc,Liu:2023cse,Liu:2024jno,Broniowski:2024oyk,Cao:2024zlf,Liu:2024rdm,Yao:2024ixu,Fujii:2024rqd,Goharipour:2025lep,Dehghan:2025ncw,Goharipour:2025yxm,Broniowski:2025ctl,Hatta:2025vhs,Hatta:2025ryj,Dehghan:2025eov,Guo:2025jiz,Liu:2025vfe,Sugimoto:2025btn,Nair:2025sfr,Cao:2025dkv,Coriano:2024wrz,Coriano:2025lge,Coriano:2025lge,Stegeman:2025sca,Fujii:2025aip,Tanaka:2025pny,Fukushima:2025jah,Tanaka:2025hhv,Stegeman:2025tdl,Xing:2025uwn,Kawaguchi:2025cuf,Fukushima:2026wwc,Ejima:2026bzo,Deng:2026ydm,Hagiwara:2024wqz} and lattice QCD simulations~\cite{Shanahan:2018pib,Shanahan:2018nnv,Pefkou:2021fni,Hackett:2023nkr,Hackett:2023rif,Wang:2024lrm}, 
and have also been discussed in connection with experimental studies~\cite{Burkert:2018bqq,Burkert:2021ith,Duran:2022xag,CLAS:2026lls}. In the low-energy region of QCD, spin-zero mesons provide a useful testing ground for studying the connection between GFFs and nonperturbative QCD dynamics, because pseudoscalar mesons, such as the pion, are closely tied to chiral symmetry breaking, while scalar mesons are particularly sensitive to chiral-partner dynamics. %reflect the corresponding chiral-partner as well as symmetry-breaking dynamics.
For a spin-zero meson $M$, the GFFs are defined as
\begin{equation}
\begin{split}
\langle M(p')| \Theta^{\mu\nu}(0) |M(p)\rangle
&= 2P^\mu P^\nu A_M(t) \\
&+ \frac{1}{2}\left(q^\mu q^\nu-q^2 g^{\mu\nu}\right)D_M(t),
\end{split}
\end{equation}
where $P=(p+p')/2$, $q=p'-p$, and $t=-q^2$. 
Here, $A_M(t)$ and $D_M(t)$ are the meson GFFs. The A-form factor encodes information on the energy-momentum distribution of the meson, and its forward ($t\to 0$) limit is fixed to 1 by energy-momentum conservation.
In contrast, the D-form factor is related to the internal mechanical structure, such as the pressure and shear-force distributions ~\cite{Polyakov:2018zvc,Burkert:2023wzr}, and may provide insight into confinement and other nonperturbative properties of QCD.
However, its forward limit is not fixed by a conserved charge. Moreover, the momentum-transfer dependence of the D-form factor, especially near the forward limit, has not yet been fully clarified.
Therefore, the physical origin and quantitative behavior of the meson D-form factor remain nontrivial problems.

Meson properties are strongly governed by the symmetry-breaking structures of QCD. 
In particular, meson masses reflect spontaneous chiral symmetry breaking, explicit chiral symmetry breaking due to current quark masses, flavor symmetry breaking associated with the difference between the light- and strange-quark masses, and the $U(1)_A$ anomaly. 
To capture these symmetry-breaking structures, the linear sigma model (LSM) provides a useful framework. 
In this model, scalar and pseudoscalar mesons are treated in a common chiral multiplet, and the relations between meson masses and symmetry-breaking interactions can be analyzed in a transparent way. 
More specifically, the conventional LSM provides a reasonable description of the pseudoscalar-meson masses, whereas it does not reproduce the observed scalar-meson mass hierarchy below $1\,{\rm GeV}$. 
In Ref.~\cite{Kuroda:2019jzm}, an interaction term was introduced to address this issue and reproduce the scalar-meson mass hierarchy. 
This term has a determinant-type structure involving the current quark mass and is associated with both the $U(1)_A$ anomaly and explicit chiral symmetry breaking. 
Although this is one specific effective treatment of symmetry-breaking interactions, the analysis illustrates that the LSM provides a useful framework for examining how such interactions are related to meson masses.

In this study, using effective-model approaches, we investigate how symmetry-breaking structures are reflected in meson GFFs, particularly in the D-form factors, in the low-energy regime of QCD. Recent studies have suggested that the lightest-scalar-meson-dominance picture provides a useful framework for describing the pion and nucleon D-form factors~\cite{Broniowski:2024oyk,Broniowski:2025ctl,Stegeman:2025sca,Stegeman:2025tdl,Kawaguchi:2025cuf}. Motivated by this picture, we first show that scalar-meson couplings connect the meson D-form factors to meson masses. We then evaluate the D-form factors of mesons below $1\,{\rm GeV}$ within the three-flavor LSM.

To examine the role of symmetry-breaking structures, we begin with the three-flavor-symmetric limit, in which analytical expressions can be derived and the effects of symmetry breaking can be analyzed transparently. We also investigate realistic parameter sets with unequal light- and strange-quark masses.

Furthermore, we compare the LSM results with those obtained from a chiral perturbation theory (ChPT) Lagrangian including a dilatonic scalar field.
This approach incorporates scale invariance and its explicit breaking.
Through this comparison, we examine how scale-symmetry-breaking properties enter the meson D-form factors, with particular attention to the canonical mass dimensions of the interaction terms that contribute to meson masses.

%%%%%%%%%%%%%%%%%%%%%%%%%%%%%%%%%%%%%%%%
\section{Isosinglet scalar meson dominance}

\begin{figure}
\begin{center}
\includegraphics[scale=0.35]{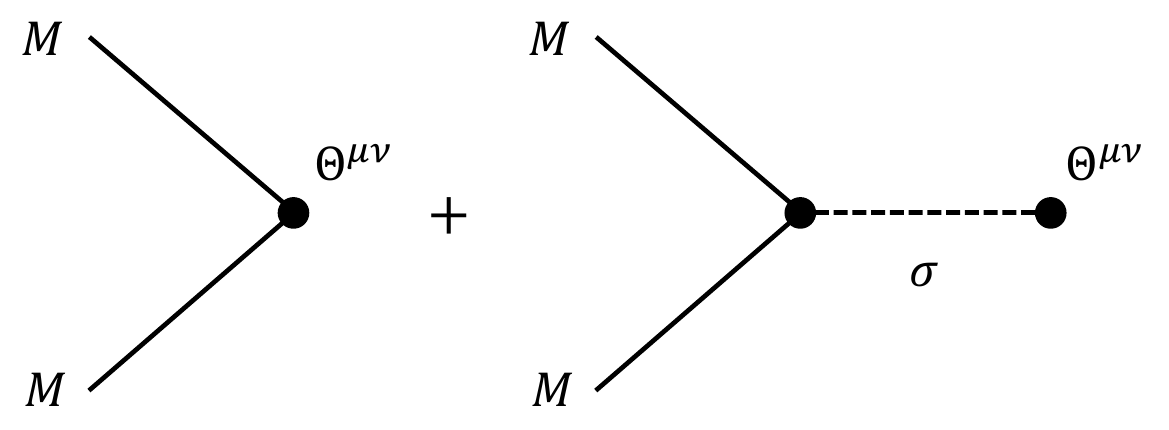}
\end{center}
\caption{Feynman diagrams contributing to the meson GFFs under the assumption of the lightest-sigma meson dominance. The first diagram represents the contact interaction $MM\Theta^{\mu\nu}$, while the second represents the contribution from the three-point $MM\sigma$ interaction mediated by the lightest scalar meson.
  }
  \label{diagram}
\end{figure}

In the low-energy region of QCD, the EMT can couple to hadronic states with the corresponding quantum numbers, such as isosinglet spin-zero scalar mesons and spin-two tensor mesons. In particular, the $f_0(500)$ meson is much lighter than other scalar mesons~\cite{ParticleDataGroup:2024cfk} and is therefore expected to play an important role in the EMT matrix elements of mesons. Recent analyses of the pion and nucleon GFFs suggest that the lightest sigma-meson dominance picture is realized in the low-energy regime~\cite{Broniowski:2024oyk,Broniowski:2025ctl,Stegeman:2025sca,Stegeman:2025tdl,Kawaguchi:2025cuf}. Motivated by this fact, in our study, we analyze the GFFs of mesons under the assumption of lightest scalar-meson dominance, as depicted in Fig.~\ref{diagram}.

Within this picture, the D-form factor of a meson $M$ may be represented schematically by
\begin{equation}
\begin{split}
D_M(t)\sim 
\mbox{(const.)}
+
\frac{
2g_{\sigma MM}\, g_{\sigma \Theta}
}{m^2_\sigma +t},
\label{rought_D}
\end{split}
\end{equation}
where $m_\sigma$ is the lightest scalar meson mass, $g_{\sigma MM}$ denotes the three-point coupling between the scalar meson and two mesons, and $g_{\sigma \Theta}$ denotes the coupling between the scalar meson and the EMT.
The contact diagram in Fig.~\ref{diagram} may be associated with the constant contribution in Eq.~\eqref{rought_D}, whereas the second diagram, arising from sigma-meson exchange through the three-point meson interaction, generates the momentum-transfer dependence.

To see how the meson properties enter the scalar-meson-exchange contribution, we explicitly consider the three-point interaction~\footnote{
In general, derivative couplings among mesons may also be present. In the present study, however, we employ a linear representation of the meson fields to focus on meson-mass generation associated with the chiral condensate. In this framework, the conventional linear sigma model provides the relevant three-point interactions through non-derivative interaction terms.
}
\begin{equation}
\begin{split}
{\cal L}_{\rm int} = -\frac{1}{2}g_{\sigma MM} \sigma M^2. 
\end{split}
\end{equation}
When the scalar meson acquires a condensate $\bar\sigma$, this interaction contributes to the mass term of the meson field.
Therefore, the coupling $g_{\sigma MM}$ can be interpreted as the response of the meson mass to the scalar condensate,
\begin{equation}
\begin{split}
g_{\sigma MM} = \frac{\partial m^2_M}{\partial \bar \sigma}.
\end{split}
\end{equation}
Indeed, meson masses are governed by several symmetry-breaking effects in QCD, such as spontaneous chiral symmetry breaking, explicit chiral symmetry breaking due to current quark masses, flavor symmetry breaking caused by the difference between the light and strange quark masses, and the $U(1)_A$ anomaly. 
Thus, within the sigma-meson-dominance picture, the D-form factor is expected to be sensitive to the symmetry-breaking structure reflected in the meson mass.
In the following sections, we examine this connection explicitly using effective model approaches.

%%%%%%%%%%%%%%%%%%%%%%%%%%%%%%%%%%%%%%%%
\section{Chiral effective model and symmetry breaking}

To describe the symmetry-breaking structure reflected in the meson mass, we employ the three-flavor linear sigma model. This model provides a convenient framework for simultaneously incorporating several symmetry-breaking structures of QCD, including the $U(1)_A$ anomaly, explicit chiral symmetry breaking, and flavor symmetry breaking. 
In this section, we briefly review the model setup and present the meson mass formulas used in the subsequent analysis of the D-form factors.
%, which will be used in the following sections to discuss how the D-form factors depend on the symmetry-breaking structures of QCD.

%%%%%%%%%%%%%%%%%%%%%%%%%%%%%%%%%%%%%%%%
\subsection{Linear sigma model}
The linear sigma model is constructed in terms of the chiral meson-field multiplet $\Phi$, which is parametrized by scalar- and pseudoscalar-meson fields as
\begin{equation}
\begin{split}
\Phi&=(S_a+i P_a)T_a,
\label{PhiField}
\end{split}
\end{equation}
where $S_a$ are the scalar fields and $P_a$ are the pseudoscalar fields.
Here, $T_a=\lambda_a/2$  $(a=0,1,\cdots,8)$ are the generators of $U(3)$ with $\lambda_{i}$ $(i=1,\cdots,8)$ being the Gell-Mann matrices and $\lambda_{a=0}=\sqrt{2/3}\, {\bm 1}_{3\times 3}$.
The scalar and pseudoscalar mesons relevant to the low-energy regime of QCD are embedded in $S_aT_a$ and $P_a T_a$, respectively:
\begin{equation}
\begin{split}
&S_a T_a\\
&=
\frac{1}{\sqrt{2}}
\begin{pmatrix}
\frac{a^0}{\sqrt{2}}+\frac{\sigma_8}{\sqrt{6}}+\frac{\sigma_0}{\sqrt{3}}&a^+&\kappa^+\\
a^-&-\frac{a^0}{\sqrt{2}}+\frac{\sigma_8}{\sqrt{6}}+\frac{\sigma_0}{\sqrt{3}} &\kappa^0\\
\kappa^-&\bar \kappa^0&-\frac{2\sigma_8}{\sqrt{6}}+\frac{\sigma_0}{\sqrt{3}}
\end{pmatrix},\nonumber\\
&P_a T_a\\
&=
\frac{1}{\sqrt{2}}
\begin{pmatrix}
\frac{\pi^0}{\sqrt{2}}+\frac{\eta_8}{\sqrt{6}}+\frac{\eta_0}{\sqrt{3}}&\pi^+&K^+\\
\pi^-&-\frac{\pi^0}{\sqrt{2}}+\frac{\eta_8}{\sqrt{6}}+\frac{\eta_0}{\sqrt{3}} &K^0\\
K^-&\bar K^0&-\frac{2\eta_8}{\sqrt{6}}+\frac{\eta_0}{\sqrt{3}}
\end{pmatrix},
\end{split}
\end{equation}
where 
$a^\pm=(S_1\mp i S_2)/{\sqrt 2}$ and $a^0 =S_3$ denote the $a_0(980)$ mesons identified as the isotriplet scalar mesons, also known as $\delta$ mesons;
$\kappa^\pm=(S_4\mp i S_5)/{\sqrt 2}$, 
$\kappa^0=(S_6- i S_7)/{\sqrt 2}$ and
$\bar \kappa^0=(S_6 + i S_7)/{\sqrt 2}$
denote the $\kappa(700)=K_0^*(700)$ mesons;
$\sigma_0=S_0$ and $\sigma_8=S_8$ are the admixtures of the $f_0(500)$ and $f_0(980)$;
$\pi^\pm=(P_1\mp i P_2)/{\sqrt 2}$ 
and $\pi^0 =P_3$ denote pions;
$K^\pm=(P_4\mp i P_5)/{\sqrt 2}$, 
$K^0=(P_6- i P_7)/{\sqrt 2}$ and
$\bar K^0=(P_6 + i P_7)/{\sqrt 2}$
denote kaons;
$\eta_0=P_0$ and $\eta_8=P_8$ are admixtures of the $\eta$ meson and the $\eta'$ meson.

Under the chiral $SU(3)_L\times SU(3)_R\times U(1)_A$ symmetry, the chiral meson field $\Phi$ transforms as 
\begin{eqnarray}
\Phi\to g_A\cdot g_L\cdot \Phi\cdot g_R^\dagger,
\end{eqnarray}
where $g_{L,R}\in SU(3)_{L,R}$ and $g_A\in U(1)_A$.
Using the building block $\Phi$, the three-flavor linear sigma model is written as
\begin{equation}
\begin{split}
{\cal L}={\rm tr}\left[ \partial_\mu \Phi\partial^\mu \Phi^\dagger \right]-V,
\end{split}
\end{equation}
where $V$ represents the potential term, decomposed as
\begin{equation}
\begin{split}
V&=V_0+V_{\rm anom}+V_{\rm SB}+V_{\rm SB-anom}.
\end{split}
\end{equation}
The term $V_0$ is the symmetric part, which is invariant under the $SU(3)_L\times SU(3)_R\times U(1)_A$ symmetry and is given by
\begin{equation}
\begin{split}
V_0=\mu^2{\rm tr }[\Phi^\dagger\Phi]
+\lambda_1{\rm tr }[(\Phi^\dagger\Phi)^2]
+\lambda_2\Bigl({\rm tr }[(\Phi^\dagger\Phi)]\Bigl)^2,
\end{split}
\end{equation}
where $\mu^2$ denotes the mass parameter and $\lambda_{1,2}$ are dimensionless quartic coupling constants.
This part contains the quadratic and quartic terms in $\Phi$, as in the standard linear sigma model, and is responsible for spontaneous chiral symmetry breaking.

The $U(1)_A$ anomalous contribution is described by the determinant term in the linear sigma model. This term preserves the chiral $SU(3)_L\times SU(3)_R$ symmetry but explicitly breaks the $U(1)_A$ symmetry. The corresponding potential term is given by
\begin{equation}
\begin{split}
V_{\rm anom}=-B\left({\rm det}[\Phi]+{\rm det}[\Phi^\dagger]\right),
\label{anomaly_int}
\end{split}
\end{equation}
where $B$ is a real coupling constant.
This term is referred to as the Kobayashi-Maskawa-'t Hooft (KMT) interaction~\cite{Kobayashi:1970ji,Kobayashi:1971qz,tHooft:1976rip,tHooft:1976snw} and generates a sufficiently large mass for the $\eta^\prime$ meson, so that it is no longer regarded as a Nambu-Goldstone boson.

The explicit chiral symmetry breaking arising from the current quark masses is introduced through the $V_{\rm SB}$ term,
\begin{equation}
\begin{split}
V_{\rm SB}=-c{\rm tr}[{\cal M}\Phi^\dagger+{\cal M}^\dagger \Phi],
\end{split}
\end{equation}
where $c$ is the real parameter and ${\cal M}$ denotes the current quark mass matrix, ${\cal M} ={\rm diag}(m_u,m_d,m_s)$.
In this study, we impose isospin symmetry, $m_u=m_d\equiv m_l$.

The term $V_{\rm SB-anom}$ was proposed in Ref.~\cite{Kuroda:2019jzm} to describe explicit chiral symmetry breaking associated with the determinant-type interaction:
\begin{equation}
\begin{split}
V_{\rm SB-anom}=-kc\left[\epsilon_{abc}\epsilon^{def}{\cal M}^a_d\Phi^b_e \Phi^c_f+{\rm h.c.}
\right],
\end{split}
\end{equation}
where $k$ denotes the real parameter and $\epsilon_{abc}$ is the totally antisymmetric tensor with $\epsilon_{123}=1$. 
This interaction term represents minimal flavor violation associated with the $U(1)_A$ anomaly and can be obtained by replacing one of the chiral fields in the $U(1)_A$ anomaly term in Eq.~\eqref{anomaly_int} with the quark mass matrix ${\cal M}$.
In Ref.~\cite{Kuroda:2019jzm}, this interaction term has been shown to play a crucial role in realizing the mass hierarchy of the scalar mesons below $1\,{\rm GeV}$: $m[f_0(500)]<m[K_0^*(700)]<m[a_0(980)]<m[f_0(980)]$.

%%%%%%%%%%%%%%%%%%%%%%%%%%%%%%%%%%%%%%%%
\subsection{Meson masses}

In the linear sigma model, spontaneous chiral symmetry breaking is realized through a nonzero vacuum expectation value of the chiral field $\Phi$. Owing to flavor symmetry breaking due to $m_l\neq m_s$, this breaking is reflected in the chiral condensate as
\begin{equation}
\begin{split}
\langle0| \Phi |0\rangle =
\bar \sigma_0 T_0 + \bar \sigma_8 T_8.
\end{split}
\end{equation}
%In the flavor-symmetric case, $m_l=m_s$, one has $\bar \sigma_8=0$, while $\bar \sigma_0$ takes a nonzero value.
Substituting these condensates into the potential $V$, we obtain the effective potential at the mean-field level. The chiral condensates are then determined from the stationary conditions for $V(\bar \sigma_0,\bar \sigma_8)$,
\begin{equation}
\begin{split}
\frac{\partial V(\bar \sigma_0, \bar \sigma_8) }{\partial \bar \sigma_0} &=0,\\
\frac{\partial V(\bar \sigma_0, \bar \sigma_8) }{\partial \bar \sigma_8} &=0.
\end{split}
\label{eq14}
\end{equation}

Around the vacuum characterized by nonzero $\bar \sigma_0$ and $\bar \sigma_8$, the mesons appear as fluctuating modes. 
The pseudoscalar meson mass matrix $(m_P^2)_{ab}$ is obtained from the second derivatives of the potential with respect to the fluctuating fields. The relevant matrix elements are given by
\begin{equation}
\begin{split}
(m_{P}^2)_{00}&=
\mu^2+
\frac{\lambda_1}{3}(\bar\sigma_0^2+\bar\sigma_8^2)
+\lambda_2(\bar\sigma_0^2+\bar\sigma_8^2)
\\
&
+B\sqrt{\frac{2}{3}}\bar\sigma_0
+\frac{4}{3}kc(2 m_l+m_s),\\
(m_{P}^2)_{88}&=
\mu^2+
\lambda_1\left[\frac{1}{3}(\bar\sigma_0)^2-\frac{\sqrt{2}}{3}\bar\sigma_0\bar\sigma_8
+\frac{1}{2}(\bar\sigma_8)^2
\right]
\\
&
+\lambda_2(\bar\sigma_0^2+\bar\sigma_8^2)
-\frac{B}{\sqrt{3}}\left(\frac{1}{\sqrt2}\bar\sigma_0+\bar\sigma_8\right)
\\
&
-\frac{2}{3}kc(4m_l-m_s),
\\
(m_{P}^2)_{08}&=
\frac{2}{3}\lambda_1\bar\sigma_0\bar\sigma_8
-\frac{\lambda_1}{3\sqrt{2}}\bar\sigma_8^2
\\
&
-\frac{B}{\sqrt{6}}\bar\sigma_8
-\frac{2\sqrt{2}}{3}kc( m_l-m_s),
\\
(m_{P}^2)_{11}&=
\mu^2+
\lambda_1\left[\frac{1}{3}\bar\sigma_0^2+\frac{\sqrt{2}}{3}\bar\sigma_0\bar\sigma_8
+\frac{1}{6}\bar\sigma_8^2\right]+
\lambda_2(\bar\sigma_0^2+\bar\sigma_8^2)
\\
&
-\frac{B}{\sqrt{3}}\left(\frac{1}{\sqrt2}\bar\sigma_0-\bar\sigma_8\right)
-2kcm_s,\\
(m_{P}^2)_{44}&=
\mu^2+
\lambda_1\left[\frac{1}{3}(\bar\sigma_0)^2-\frac{1}{3\sqrt{2}}\bar\sigma_0\bar\sigma_8
+\frac{7}{6}(\bar\sigma_8)^2\right]\\
&
+\lambda_2(\bar\sigma_0^2+\bar\sigma_8^2)
-\frac{B}{\sqrt{3}}\left(\frac{1}{\sqrt2}\bar\sigma_0+\frac{1}{2}\bar\sigma_8\right)
-2kc m_l.
\end{split}
\end{equation}
The components $(m_{P}^2)_{11}$ and $(m_{P}^2)_{44}$ are identified with the pion and kaon masses, respectively: $m_\pi^2\equiv (m_{P}^2)_{11}$,  $m_K^2\equiv (m_{P}^2)_{44}$.
In the chiral limit, the masses of these pseudoscalar mesons vanish after imposing the stationary conditions, indicating that they are Nambu-Goldstone (NG) bosons.

For the singlet and octet components,
the diagonal elements $(m_P^2)_{00}$ and $(m_P^2)_{88}$ are associated with the $\eta_0$ and $\eta_8$ sectors, respectively: $m_{\eta_0}^2\equiv (m_P^2)_{00}$ and $m_{\eta_8}^2\equiv (m_P^2)_{88}$.
Owing to flavor symmetry breaking for $m_l\neq m_s$, mixing occurs through the off-diagonal element, $m_{P{\rm mix}}^2\equiv (m_P^2)_{08}$. We therefore introduce the mass eigenstates by rotating the singlet and octet fields with the mixing angle $\theta_P$ as
\begin{equation}
\begin{split}
\begin{pmatrix}
\tilde \pi_1 \\
\tilde \pi_2
\end{pmatrix}
&=
O_P^T
\begin{pmatrix}
P_0 \\
P_8
\end{pmatrix},
\end{split}
\end{equation}
with
\begin{equation}
O_P = 
\begin{pmatrix}
  \cos\theta_P & -\sin \theta_P \\
 \sin \theta_P  & \cos \theta_P
\end{pmatrix}.
\end{equation}
By diagonalizing the mass matrix with this rotation, the mass eigenvalues are identified with the $\eta'$ and $\eta$ mesons as
\begin{equation}
\begin{split}
\begin{pmatrix}
m_{\eta'}^2 &0\\
0& m_\eta^2
\end{pmatrix}
&=
O_P^T 
\begin{pmatrix}
m_{\eta_0}^2 &
m_{P{\rm mix}}^2
\\
m_{P{\rm mix}}^2
& m_{\eta_8}^2
\end{pmatrix}
O_P. 
\end{split}
\end{equation}
In the flavor-symmetric limit, this mixing vanishes, $m_{P{\rm mix}}^2=0$. Furthermore, in the chiral limit, the pseudoscalar meson associated with the octet component becomes a NG boson. On the other hand, the singlet component remains massive even in the chiral limit because of the presence of the $U(1)_A$ anomalous interaction in Eq.~\eqref{anomaly_int}.

Similarly to the pseudoscalar meson sector, the elements of the scalar meson mass matrix $(m_{S}^2)_{ab}$ take the following form:
\begin{equation}
\begin{split}
(m_{S}^2)_{00}&=
\mu^2+\lambda_1(\bar\sigma_0^2+\bar\sigma_8^2)
+\lambda_2(3\bar\sigma_0^2+\bar\sigma_8^2)\\
&
-B\sqrt{\frac{2}{3}}\bar\sigma_0
-\frac{4}{3}kc(2m_l+m_s),\\
(m_{S}^2)_{88}&=
\mu^2+
\lambda_1\left[\bar\sigma_0^2-\sqrt{2}\bar\sigma_0\bar\sigma_8
+\frac{3}{2}\bar\sigma_8^2
\right]+
\lambda_2(\bar\sigma_0^2+3\bar\sigma_8^2)
\\
&
+\frac{B}{\sqrt{3}}\left(\frac{1}{\sqrt2}\bar\sigma_0+\bar\sigma_8\right)
+\frac{2}{3}kc(4 m_l-m_s),
\\
(m_{S}^2)_{08}&=
\lambda_1\left[2\bar\sigma_0\bar\sigma_8
-\frac{1}{\sqrt{2}}\bar\sigma_8^2\right]
+2\lambda_2\bar\sigma_0\bar\sigma_8
+\frac{B}{\sqrt{6}}\bar\sigma_8
\\
&
+\frac{2\sqrt{2}}{3}kc( m_l-m_s),
\\ 
(m_{S}^2)_{11}&=
\mu^2+
\lambda_1\left[\bar\sigma_0^2+\sqrt{2}\bar\sigma_0\bar\sigma_8
+\frac{1}{2}\bar\sigma_8^2\right]+
\lambda_2(\bar\sigma_0^2+\bar\sigma_8^2)
\\
&
+\frac{B}{\sqrt{3}}\left(\frac{1}{\sqrt2}\bar\sigma_0-\bar\sigma_8\right)
+2kcm_s,
\\
(m_{S}^2)_{44}&=\mu^2+
\lambda_1\left[\bar\sigma_0^2-\frac{1}{\sqrt{2}}\bar\sigma_0\bar\sigma_8
+\frac{1}{2}\bar\sigma_8^2\right]+
\lambda_2(\bar\sigma_0^2+\bar\sigma_8^2)
\\
&
+\frac{B}{\sqrt{3}}\left(\frac{1}{\sqrt2}\bar\sigma_0+\frac{1}{2}\bar\sigma_8\right)
+2kc m_l.
\end{split}
\end{equation}
The $(m_{S}^2)_{11}$ and $(m_{S}^2)_{44}$ are identified with the masses of the $a_0(980)$ and $K_0^*(700)$ mesons, respectively:
$m^2_\delta\equiv (m_{S}^2)_{11}$ and $m^2_\kappa\equiv (m_{S}^2)_{44}$.
In the present linear sigma model,  $m_\delta^2$ contains the contribution $2kcm_s$, whereas $m_\kappa^2$ contains $2kcm_l$. Since $m_s\gg m_l$, the term $V_{\rm SB-anom}$ generates a sufficiently large mass splitting, making the $a_0(980)$ heavier than the $K_0^*(700)$.

Mixing also occurs in the scalar meson sector owing to flavor symmetry breaking. 
The diagonal elements $(m_S^2)_{00}$ and $(m_S^2)_{88}$ are associated with the $\sigma_0$ and $\sigma_8$ sectors, respectively: $m_{\sigma_0}^2\equiv(m_S^2)_{00}$ and $m_{\sigma_8}^2\equiv(m_S^2)_{88}$.
The mixing arises from the off-diagonal element, $m_{S{\rm mix}}^2\equiv(m_S^2)_{08}$.
Therefore, we define the mass eigenstates by rotating the singlet and octet scalar fields with the mixing angle $\theta_S$:
\begin{equation}
\begin{split}
\begin{pmatrix}
\phi_1 \\
 \phi_2
\end{pmatrix}
&=
O_S^T
\begin{pmatrix}
S_0 \\
S_8
\end{pmatrix},
\label{scalar_es}
\end{split}
\end{equation}
with
\begin{equation}
O_S = 
\begin{pmatrix}
  \cos\theta_S & -\sin \theta_S \\
 \sin \theta_S  & \cos \theta_S
\end{pmatrix}.
\end{equation}
The diagonalized scalar meson masses are identified with the masses of the $f_0(500)$ and $f_0(980)$ mesons as
\begin{equation}
\begin{split}
\begin{pmatrix}
m_{f_0(500)}^2 &0\\
0& m_{f_0(980)}^2
\end{pmatrix}
&=
O_S^T 
\begin{pmatrix}
m_{\sigma_0}^2 &
m_{S{\rm mix}}^2
\\
m_{S{\rm mix}}^2
& m_{\sigma_8}^2
\end{pmatrix}
O_S. 
\end{split}
\end{equation}

%%%%%%%%%%%%%%%%%%%%%%%%%%%%%%%%%%%%%%%%
\section{Meson D-form factors in LSM}

In this section, we evaluate the meson D-form factors within the LSM.
We first determine the coupling between the scalar meson and the EMT.
Then, following the scalar-meson dominance picture, we express the meson D-form factors.
To investigate their structure in more detail, we consider both the three-flavor symmetric limit and the realistic case with different light- and strange-quark masses.

%%%%%%%%%%%%%%%%%%%%%%%%%%%%%%%%%%%%%%%%
\subsection{EMT in LSM}
\label{EMT_LSM_subsec}

In the linear sigma model, the energy-momentum tensor (EMT) takes the form
\begin{equation}
\begin{split}
\Theta^{\mu\nu} &=
2  {\rm tr}\left[ \partial^\mu \Phi\partial^\nu \Phi^\dagger \right]
-g^{\mu \nu}{\cal L}
\\
&
-\frac{1}{3} (\partial_\mu \partial_\nu - g_{\mu\nu} \partial^2  ) {\rm tr}[\Phi \Phi^\dagger].
\label{EMT_LSM}
\end{split}
\end{equation}
The last term in Eq.~\eqref{EMT_LSM} is the improvement term~\cite{Callan:1970ze}. This term is introduced so that the trace of the EMT is properly related to the divergence of the dilatation current, ${\Theta^{\mu}}_\mu =\partial_\mu j^\mu_D$,
with $j^\mu_D=\Theta^{\mu\nu}x_\nu$.

After expanding the chiral field around the vacuum determined by Eq.~(\ref{eq14}), the improvement term contains contributions proportional to the chiral condensates. As a result, the EMT couples to the isosinglet scalar mesons. The corresponding matrix element is given by
\begin{equation}
\begin{split}
\langle 0| 
\Theta^{\mu\nu}(x) 
| \phi_{1(2)} (p)\rangle 
&=
\frac{\bar \phi_{1(2)}}{3} 
(p^\mu p^\nu -g^{\mu\nu}  p^2  ) e^{-ip\cdot x},
\label{EMT_sigma}
\end{split}
\end{equation}
where $\phi_1$ and $\phi_2$ denote the mass eigenstates of the scalar mesons as shown in Eq.~\eqref{scalar_es}, which are identified with the $f_0(500)$ and $f_0(980)$ mesons, respectively. The quantities $\bar\phi_1$ and $\bar\phi_2$ are defined by rotating the condensates in the singlet-octet basis as
\begin{equation}
\begin{split}
\begin{pmatrix}
\bar \phi_1 \\
\bar \phi_2
\end{pmatrix}
&=
O_S^T
\begin{pmatrix}
\bar \sigma_0 \\
\bar \sigma_8
\end{pmatrix}.
\end{split}
\end{equation}
The coupling constants between the EMT and the scalar mesons can therefore be read off from Eq.~\eqref{EMT_sigma} as
\begin{equation}
\begin{split}
g_{\phi_{1(2)} \Theta}=
-\frac{\bar \phi_{1(2)}}{3}.
\end{split}
\end{equation}

Note that the coupling between the sigma meson and the EMT can be naturally formulated within a dilaton framework. Nevertheless, there may be a connection between the LSM and the dilaton framework. When the nonlinear representation is adopted for the linear field multiplet $\Phi$, the radial component associated with the lightest scalar mode is a chiral singlet, suggesting a possible interpretation as a dilatonic scalar. In this sense, the sigma meson in the LSM may be connected to a dilatonic degree of freedom. We do not pursue this issue further in the present study, but use the corresponding coupling as an effective input in the following analysis of the D-form factor.

%%%%%%%%%%%%%%%%%%%%%%%%%%%%%%%%%%%%%%%%
\subsection{
Scalar meson dominance in LSM}
To derive the meson D-form factor, we focus on the relevant three-point meson interaction terms,
\begin{equation}
\begin{split}
{\cal L}_{\sigma MM}
&=-\frac{1}{2}g_{\phi_1 MM}\phi_1 M^2
-\frac{1}{2}g_{\phi_2 MM}\phi_2 M^2,
\end{split}
\end{equation}
where $M$ collectively denotes one of the meson mass eigenstate fields considered in this study, and the coupling constants $g_{\phi_{1,2} MM}$ are given by
\begin{equation}
\begin{split}
g_{\phi_1 MM}&=\frac{\partial m_{M}^2}{\partial \bar\sigma_0}\cos \theta_S
+
\frac{\partial m_{M}^2}{\partial \bar\sigma_8}\sin \theta_S
,\\
g_{\phi_2 MM}&=-\frac{\partial m_{M}^2}{\partial \bar\sigma_0}\sin \theta_S
+
\frac{\partial m_{M}^2}{\partial \bar\sigma_8}\cos \theta_S.
\end{split}
\end{equation}
Note that for the three-point self-interactions of $\phi_1$ and $\phi_2$, the normalization factor is different. In this case, the corresponding interaction terms are written as
\begin{equation}
\begin{split}
{\cal L}_{\sigma MM}
&=-\frac{1}{6}g_{\phi_1 \phi_1\phi_1}\phi_1^3
-\frac{1}{6}g_{\phi_2 \phi_2\phi_2}\phi_2^3.
\end{split}
\end{equation}
%The explicit expressions for the three-point couplings are given in Appendix~\ref{three_int}.

Using these meson interaction terms within the isosinglet scalar-meson dominance picture in Fig.~\ref{diagram}, we obtain the meson D-form factor as
\begin{equation}
\begin{split}
D_{M}(t)&=
-1
+\frac{2}{3} %-\frac{2}{3}
+
\left(
\frac{2 g_{\phi_1 MM}
g_{\phi_{1} \Theta} }{t+ m_{f_0(500)}^2}
+ 
\frac{2g_{\phi_2 MM} g_{\phi_{2} \Theta}  }{t+  m_{f_0(980)}^2}
\right)
.
\label{LSM_DGFF}
\end{split}
\end{equation}
The first two terms are universal contributions originating from the contact term, whereas the last term encodes the scalar-meson exchange contribution through the three-point interaction terms. In contrast to the rough estimate in Eq.~\eqref{rought_D}, the exchange of the heavier scalar meson, $f_0(980)$, also contributes to the D-form factor. This contribution arises from the mass mixing between $\sigma_0$ and $\sigma_8$. Indeed, in the flavor-symmetric limit, where this mixing is absent, the heavier scalar-meson-exchange contribution vanishes.

The analytical expression for the D-form factor in Eq.~\eqref{LSM_DGFF} is somewhat complicated by the mass mixing induced by flavor symmetry breaking, which obscures the dependence on the underlying symmetry-breaking effects. To make the roles of explicit chiral symmetry breaking and the $U(1)_A$ anomaly more transparent, we consider the flavor-symmetric limit in the following subsection.

%%%%%%%%%%%%%%%%%%%%%%%%%%%%%%%%%%%%%%%%
\subsection{D-form factors in the flavor symmetric limit}
In this subsection, we consider the flavor-symmetric limit, $m_l=m_s$. In this limit, the octet condensate vanishes, $\bar\sigma_8=0$, and only the isosinglet condensate $\bar\sigma_0$ takes a nonzero value. Accordingly, the singlet-octet mixing in both the scalar and pseudoscalar meson sectors disappears, and the corresponding meson masses reduce to
\begin{equation}
\begin{split}
m_{\eta'}^2 &= m_{\eta_0}^2,\;\;\;
m_{\eta}^2 = m_{\eta_8}^2,\nonumber\\
m_{f_0(500)}^2 &= m_{\sigma_0}^2,\;\;\;
m_{f_0(980)}^2 = m_{\sigma_8}^2.
\end{split}
\end{equation}
Moreover, the scalar-meson masses as well as pseudoscalar-meson masses become degenerate within each $SU(3)$ flavor octet:
\begin{equation}
\begin{split}
m_{\pi}^2 &= m_{K}^2= m_{\eta_8}^2,\\
m_{\delta}^2&=m_{\kappa}^2= m_{\sigma_8}^2.
\end{split}
\end{equation}
In particular, the mass difference between the isosinglet and octet pseudoscalar mesons arises from the $U(1)_A$ anomaly terms in $V_{\rm anom}$ and $V_{\rm SB-anom}$:
\begin{equation}
\begin{split}
m_{\eta_0}^2-m_{\pi}^2 &= 
\sqrt{\frac{3}{2}}\bar \sigma_0 B + 6kcm_l.
\end{split}
\end{equation}

Under this condition, the D-form factors of the octet pseudoscalar  mesons take the common form given by
\begin{equation}
\begin{split}
D_{\pi,K,\eta_8}^{SU(3)_f }
(t)
&=
-1
+\frac{2}{3}
\frac{t + m_\pi^2  - 2kcm_l}{t+ m_{\sigma_0}^2},
\label{3fl_Dpi}
\end{split}
\end{equation}
where the superscript $SU(3)_f$ indicates that the three-flavor symmetric limit has been taken. Interestingly, this D-form factor is expressed in terms of the pion mass $m_\pi$ and the singlet scalar-meson mass $m_{\sigma_0}$. Furthermore, it contains the parameter combination $k c m_l$, which originates from $V_{\rm SB-anom}$.

When taking the chiral limit, the pion mass and $kcm_l$ vanish in Eq.~\eqref{3fl_Dpi}. As a result, the D-term in the forward limit becomes independent of meson masses and model parameters:
\begin{equation}
\begin{split}
&
\lim_{m_{l,s}=0}
D_{\pi,K,\eta_8}^{SU(3)_f }
(0) =-1.
\end{split}
\end{equation}
This result is consistent with the soft-pion theorem and previous ChPT analysis~\cite{Donoghue:1991qv}.

We next turn to the D-form factor of $\eta_0$. The $U(1)_A$ anomaly contribution induces a deviation from the D-form factor of the octet pseudoscalar mesons, yielding
\begin{equation}
\begin{split}
D_{\eta_0}^{SU(3)_f }(t)&=
D_\pi^{SU(3)_f }(t)
%+\frac{2}{9}
-\frac{2}{3}
\frac{
(m_{\eta_0}^2 -m_\pi^2)
-6kcm_l
}{t+ m_{\sigma_0}^2}.
\label{D_eta0}
\end{split}
\end{equation}
The second term on the right-hand side is characterized by the $U(1)_A$ anomaly contribution from $V_{\rm anom}$ and by the explicit chiral symmetry breaking effects associated with $V_{\rm SB}$ and $V_{\rm SB-anom}$.
Since $m_{\eta_0}^2$ remains finite even in the chiral limit owing to the $U(1)_A$ anomaly, the forward-limit value is given by
\begin{equation}
\begin{split}
\lim_{m_{l,s}=0}
D_{\eta_0}^{SU(3)_f }(0)&=
-1-\frac{2}{3}
\frac{
m_{\eta_0}^2 
}{m_{\sigma_0}^2}.
\label{Deta0}
\end{split}
\end{equation}
This shows that the D-term of the $\eta_0$ meson is reduced compared to that of the pion due to the nonzero $m_{\eta_0}^2$. 
Notably, the $\eta_0$ mass is connected to the topological susceptibility in pure Yang-Mills theory through the Witten--Veneziano relation~\cite{Witten:1979vv,Veneziano:1979ec}, $\chi_{\rm top}^{\rm YM}=
(f_\pi^2/(2N_f) )m_{\eta_0}^2$, 
where $\chi_{\rm top}^{\rm YM}$ denotes the topological susceptibility defined through the two-point correlation function of the topological charge density.
Thus, the additional negative contribution to the D-term of the $\eta_0$ meson can be attributed to the nonzero $\chi_{\rm top}^{\rm YM}$.

We now move on to the scalar-meson sector.
In the $SU(3)_f$ symmetric limit, the common D-form factor of the scalar-octet mesons is given by
\begin{equation}
\begin{split}
D_{\delta,\kappa,\sigma_8}^{SU(3)_f }(t)
&=
D_\pi^{SU(3)_f }(t)
\\
&
+\frac{4}{9}
\frac{
(m_{\eta_0}^2
-m_{\delta}^2)
+2(m_{\pi}^2
-m_{\delta}^2)  
+6kcm_l}{t+ m_{\sigma_0}^2}.
\label{D_SO_t}
\end{split}
\end{equation}
As in the case of $\eta_0$, the octet-scalar meson D-form factor can also be expressed in terms of the pion D-form factor with an additional correction term.
This correction is governed by the mass splittings associated with the chiral-partner pair ($\eta_0$, $\delta$) and the $U(1)_A$-partner pair of ($\pi$, $\delta$), together with the effect of $V_{\rm SB-anom}$.

For the isosinglet scalar meson, the corresponding D-form factor can also be expressed in terms of the pion D-form factor with an additional term, as follows:
\begin{equation}
\begin{split}
D_{\sigma_0}^{SU(3)_f }(t)
&=
D_\pi^{SU(3)_f }(t)
\\
&
+\frac{2}{9}
\frac{
6(m^2_{\pi} -m^2_{\sigma_0})
+ (m^2_{\pi}
-
m^2_{\eta_0})
-6kcm_l}{t+ m^2_{\sigma_0}},
\label{D_SI_t}
\end{split}
\end{equation}
As in the case of the scalar-octet mesons, the additional term is determined by the meson mass splittings associated with the chiral and $U(1)_A$ transformations, as well as by the contribution from $V_{\rm SB-anom}$.

In the chiral limit, the D-terms in the scalar-meson sector are given by
\begin{equation}
\begin{split}
\lim_{m_{l,s}=0}
D_{\delta,\kappa,\sigma_8}^{SU(3)_f }
(0)
&=
-1
+\frac{4}{9}
\frac{
m_{\eta_0}^2
-3m_{\delta}^2
}{m_{\sigma_0}^2}
,\\
\lim_{m_{l,s}=0}
D_{\sigma_0}^{SU(3)_f }(0)
%&=-1-\frac{2}{9}\frac{6m^2_{\sigma_0}+m^2_{\eta_0}}{m^2_{\sigma_0}}\\
&=-\frac{7}{3}
-\frac{2}{9}\frac{m^2_{\eta_0}}{m^2_{\sigma_0}}.
\end{split}
\end{equation}
This shows that the scalar-meson D-terms are also sensitive to the $U(1)_A$-anomaly contribution, which is related to the topological quantity $\chi_{\rm top}^{\rm YM}$.

%%%%%%%%%%%%%%%%%%%%%%%%%%%%%%%%%%%%%%%%
\subsection{D-form factors in realistic flavor symmetry breaking}
We now return to the more realistic case with $m_l \neq m_s$ and numerically evaluate the momentum-transfer dependence of the meson D-form factors based on Eq.~\eqref{LSM_DGFF}.

For the numerical analysis, we employ the two parameter sets listed in Table~\ref{parameter_sets}.
Both parameter sets give similar values for the pseudoscalar-meson masses, whereas they lead to different scalar-meson masses.
Set I corresponds to the conventional parameter set of the linear sigma model, in which the parameter $k$ is set to zero.
With this choice, the observed mass hierarchy of the scalar mesons,
$m[f_0(500)] < m[K_0^*(700)] < m[a_0(980)] < m[f_0(980)]$,
is not reproduced, as shown in Table~\ref{meson_masses}.
In contrast, in Set II, a nonzero value of the parameter $k$ is introduced so as to reproduce this mass hierarchy.
By comparing the results obtained with these two parameter sets, we examine how the parameter $k$ affects the meson D-form factors.

\begin{widetext}

\begin{table}[t]
\centering
\begin{tabular}{c||ccccccc}
\hline
 & $\mu^2$ & $\lambda_1$ & $\lambda_2$ & $cm_l$ & $cm_s$&$B$&$k$
 %&$\theta_P$& $\theta_S$
 \\
 & $[10^4\,{\rm MeV}^2]$
 & -
 & -
 & $[10^5\,{\rm MeV}^3]$
 & $[10^5\,{\rm MeV}^3]$
 & $[10^3\,{\rm MeV}]$
 & $[{\rm GeV}^{-1}]$ 
 %&-&-
 \\
\hline
Set I  & $11.7$ & $46.5$ & $1.40$ & $8.80$ & $269$ & $4.81$ & $0$ 
%&$-5^\circ$&$19.9^\circ$
\\
Set II & $1.02$ & $11.8$ & $20.4$ & $6.11$ & $198$ & $3.85$ & $3.40$ 
%&$6.8^\circ$& $44.3^\circ$
\\
\hline
\end{tabular}
\caption{
Input parameter sets used to calculate the meson masses. Set I, taken from Ref.~\cite{Lenaghan:2000ey}, excludes $V_{\rm SB-anom}$, while Set II, taken from Ref.~\cite{Kuroda:2019jzm}, includes it  with the nonzero value of $k$.
}
\label{parameter_sets}
\end{table}

\begin{table}[t]
\centering
\begin{tabular}{c|cccccccc|cc}
\hline
 & $m_{f_0(500)}$ & $m_{f_0(980)}$ & $m_{\delta}$ & $m_{\kappa}$ & $m_{\eta'}$ & $m_{\eta}$ & $m_{\pi}$ & $m_K$ 
 &$\theta_P$& $\theta_S$
 \\
\hline
Set I  & $600$ & $1221$ & $1028$ & $1124$ & $963$ & $539$ & $138$ & $496$ 
&$-5^\circ$
&$19.9^\circ$
\\
Set II & $672$ & $990$ & $938$ & $863$ & $958$ & $553$ &
$138$ & $494$
&
$6.8^\circ$& $44.3^\circ$
\\
\hline
\end{tabular}
\caption{
Meson masses and mixing angles obtained using the parameter sets listed in Table~\ref{parameter_sets}. All masses are given in MeV.
}
\label{meson_masses}
\end{table}

\end{widetext}

\begin{figure}[h]
\centering
\includegraphics[width=0.42\textwidth]{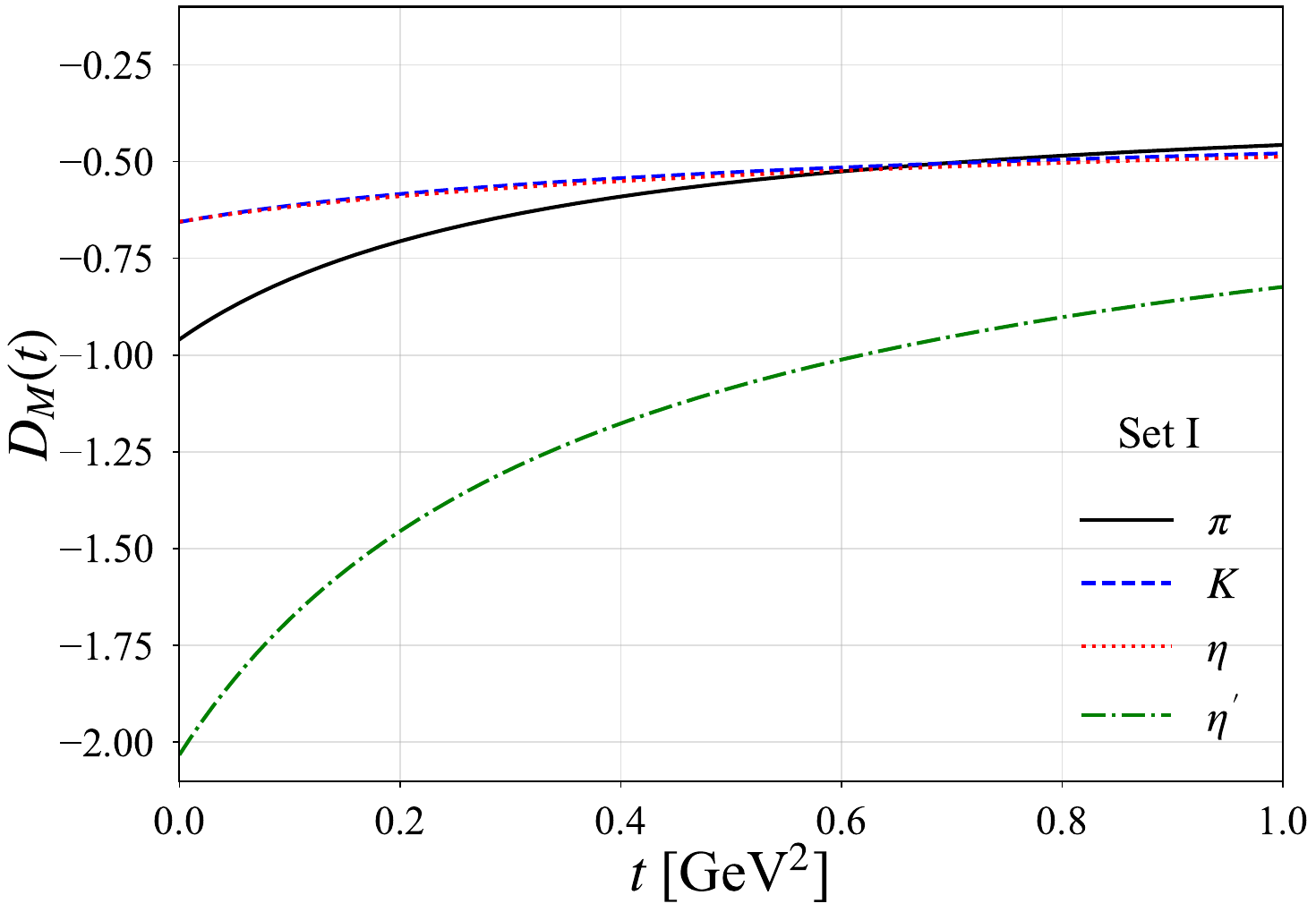}\\
(a) 

\vspace{5mm}
\includegraphics[width=0.42\textwidth]{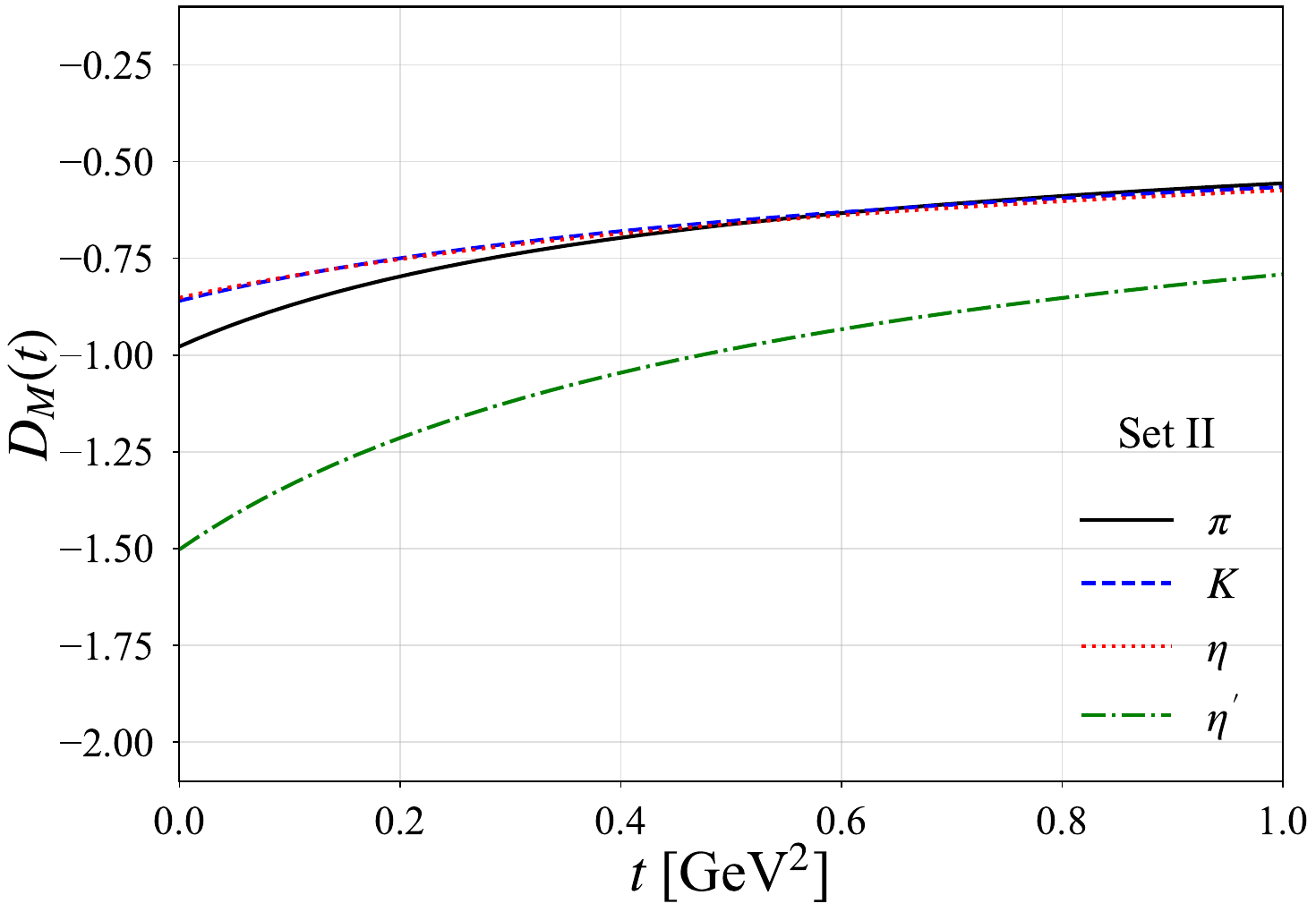}\\
(b)
\caption{
Momentum-transfer dependence of the D-form factors of the pseudoscalar mesons for (a) Set I and (b) Set II.
}
\label{Dform_LSM_PS}
\end{figure}

We first show the results for the D-form factors of the pseudoscalar mesons in Fig.~\ref{Dform_LSM_PS}.
The qualitative behavior is similar for Set I and Set II.
In the forward limit, $t=0$, the pion D-term takes a value slightly larger than $-1$.
The kaon and $\eta$-meson D-terms are almost identical to each other and are larger than the pion D-term.
On the other hand, the $\eta'$-meson D-term is smaller than the pion D-term. 
This tendency may be qualitatively understood from the $U(1)_A$ anomaly contribution in the three-flavor symmetric limit, as shown in Eq.~\eqref{Deta0}.
Furthermore, this decrease is further reduced when a nonzero value of the parameter $k$ is introduced. 
This behavior would also be consistent with the three-flavor symmetric result in Eq.~\eqref{D_eta0}, although the flavor mixing effects are sizable in the present parameter sets.

Regarding the momentum-transfer dependence, the D-form factors of the pseudoscalar mesons increase with increasing $t$. In the large-$t$ region, the scalar-meson-exchange contribution becomes less significant, and all the pseudoscalar-meson D-form factors are expected to approach $-1/3$, as follows from the analytical expression in Eq.~\eqref{LSM_DGFF}. However, this asymptotic behavior is not visible within the range of momentum transfer shown in Fig.~\ref{Dform_LSM_PS}.

\begin{figure}[h]
\centering
\includegraphics[width=0.42\textwidth]{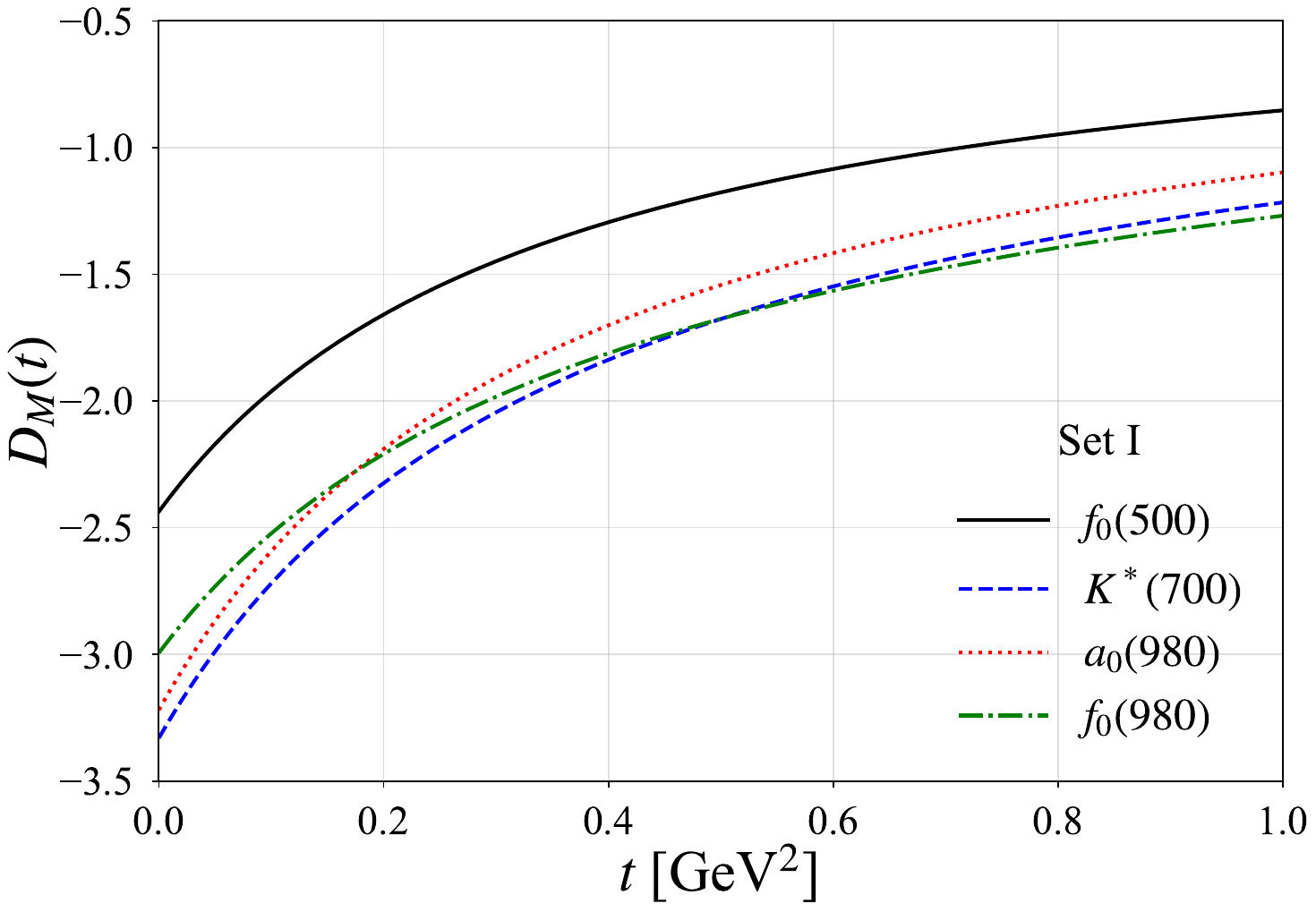}\\
(a) 

\vspace{5mm}
\includegraphics[width=0.42\textwidth]{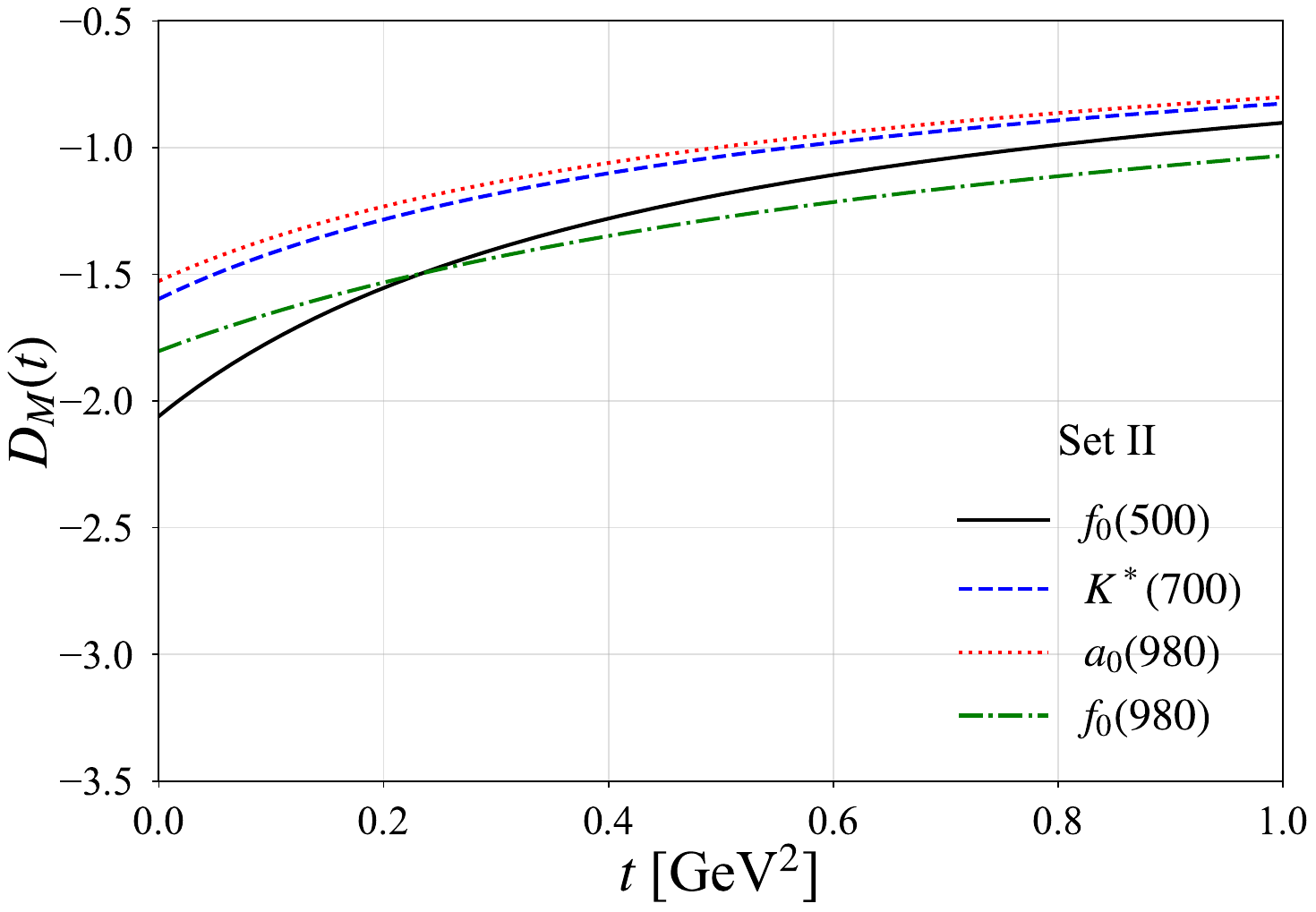}\\
(b)
\caption{
Same as Fig.~\ref{Dform_LSM_PS}, but for the scalar mesons.
}
\label{Dform_LSM_S}
\end{figure}

In Fig.~\ref{Dform_LSM_S}, we show the results for the D-form factors of the scalar mesons.
In contrast to the pseudoscalar sector, the dependence on the parameter $k$ is clearly visible in the scalar-meson D-form factors.
For Set I, where the parameter $k$ is not introduced, the scalar-meson D-terms in the forward limit are ordered as
\begin{equation}
D_{K_0^*(700)}(0) < D_{a_0(980)}(0) < D_{f_0(980)}(0) < D_{f_0(500)}(0) .
\end{equation}
On the other hand, when the parameter $k$ is introduced in Set II, this ordering changes to
\begin{equation}
D_{f_0(500)}(0) < D_{f_0(980)}(0) < D_{K_0^*(700)}(0) < D_{a_0(980)}(0) .
\end{equation}

The most notable change is found in the ordering of the D-term of  $f_0(500)$.
This behavior can be qualitatively understood from the analytical expressions in the flavor-symmetric limit, Eqs.~(\ref{D_SO_t}) and (\ref{D_SI_t}).
In this limit, the $k$-dependent contribution enters the scalar-octet D-terms with a positive sign, whereas it enters the D-term of $f_0(500)$ with a negative sign.
Therefore, a nonzero value of $k$ tends to lower the D-term of $f_0(500)$ relative to the other scalar-meson D-terms.
For the other scalar mesons, however, the sizable singlet-octet scalar-meson mixing also plays an important role.
Thus, their detailed ordering cannot be understood from the flavor-symmetric expressions alone, but should be interpreted from the full mixed result.

The momentum-transfer dependence also differs between the two parameter sets.
This behavior may also be understood as a consequence of both $V_{\rm SB-anom}$ and the singlet-octet scalar-meson mixing.

%%%%%%
%%%%%%
%%%%%%
\begin{comment}
As the momentum transfer $t$ increases, these orderings are further modified.
For Set I, at $t=1\,{\rm GeV}^2$, the ordering becomes
\begin{equation}
D_{f_0(980)}(t) < D_{K_0^*(700)}(t) < D_{a_0(980)}(t) < D_{f_0(500)}(t),
\end{equation}
whereas for Set II it becomes
\begin{equation}
D_{f_0(980)}(t) < D_{f_0(500)}(t) < D_{K_0^*(700)}(t) < D_{a_0(980)}(t) .
\end{equation}
Therefore, the scalar-meson D-form factors are sensitive to the explicit chiral-symmetry breaking associated with the determinant-type interaction $V_{\rm SB-anom}$, suggesting a connection with the mass hierarchy of scalar mesons below $1\,{\rm GeV}$.
\end{comment}
%%%%%%
%%%%%%
%%%%%%

%%%%%%%%%%%%%%%%%%%%%%%%%%%%%%%%%%%%%
\section{
Role of Scale Symmetry Breaking in the D-form Factor
}

So far, we have investigated the meson D-form factors within the LSM. In this section, we adopt a different effective-model approach based on a ChPT Lagrangian including a dilatonic scalar field and examine how scale symmetry breaking enters the meson D-form factors. In particular, we focus on the canonical mass dimensions of the interaction terms, which determine their behavior under scale transformations. To make these effects more transparent, we work in the chiral limit.

%%%%%%%%%%%%%%%%%%%%%%%%%%%%%%%%%%%%%
\subsection{ChPT with dilaton}
As a representative example of a different effective approach, we consider the three-flavor ChPT Lagrangian including a dilatonic scalar field in the chiral limit (see Refs.~\cite{Lanik:1984fc,Ellis:1984jv,Leung:1989hw,Campbell:1990ak,Donoghue:1991qv,Brown:1991kk,Song:1997kx,Lee:2003eg,Park:2003sd,Park:2008zg,Crewther:2013vea,Matsuzaki:2013eva,Li:2016uzn,Hansen:2016fri,Appelquist:2017wcg,Appelquist:2017vyy,Cata:2019edh,Appelquist:2019lgk,Brown:2019ipr,Zwicky:2023bzk,Zwicky:2023krx,Shifman:2023jqn} for related developments in dilaton effective theories):
\begin{equation}
\begin{split}
{\cal L} &=
\frac{f_\pi^2}{4}
\left(\frac{\chi}{f_\sigma}\right)^2
{\rm tr}\left[
\partial_\mu U \partial^\mu U^\dagger
\right]
+\frac{1}{2}\partial_\mu \chi \partial^\mu\chi
\\
&-V_{\rm anom}^{\rm (ChPT)}
-V_{\sigma}^{\rm (ChPT)}
%\\
%&+\tau\left(\frac{\chi}{f_\sigma}\right)^{4+\alpha}\left({\rm det } U +{\rm det } U^\dagger -2\right)\\
%&-\frac{1}{4}f_\sigma^2 m_\sigma^2\left(\frac{\chi}{f_\sigma}\right)^4\left[\ln \left(\frac{\chi}{f_\sigma}\right)-\frac{1}{4}\right]
\end{split}
\end{equation}
Here, $U$ is the chiral field in the nonlinear representation, defined as $U= \exp[i\sqrt{2/3}{\eta_0}/{f_\pi}
+i{2\pi^iT^i}/{f_\pi}
]$.
The field $\chi$ is the conformal compensator including the dilatonic scalar field. We parametrize it as
$\chi=f_\sigma + \sigma$, where
$\sigma$ is the lightest scalar meson field and
$f_\sigma$ is its decay constant.  
The $U(1)_A$ anomalous potential and the $\sigma$ potential are given by 
\begin{equation}
\begin{split}
V_{\rm anom}^{\rm (ChPT)}&=-\tau\left(\frac{\chi}{f_\sigma}\right)^{4+\alpha}
\left(
{\rm det } U +
{\rm det } U^\dagger -2
\right),\\
V_{\sigma}^{\rm (ChPT)}
&=\frac{1}{4}f_\sigma^2 m_\sigma^2
\left(\frac{\chi}{f_\sigma}\right)^4\left[
\ln \left(\frac{\chi}{f_\sigma}\right)
-\frac{1}{4}
\right],
\end{split}
\end{equation}
where $m_\sigma$ is the mass of the lightest scalar field, which is identified with $m_{\sigma_0}$ in the LSM model.
In $V_{\rm anom}^{\rm (ChPT)}$, we introduce the parameter $\alpha$, which parametrizes a possible anomalous contribution to the scaling dimension of this term.
If the standard construction of scale-invariant effective theories is followed, the $U(1)_A$-anomaly term carries the fourth power of $\chi/f_\sigma$, corresponding to $\alpha=0$.
Here, however, we keep $\alpha$ as a free parameter and examine its effect on the D-form factors.
The potential $V_{\sigma}^{\rm (ChPT)}$ is chosen to be a Coleman-Weinberg-type logarithmic potential, as employed in Refs.~\cite{Matsuzaki:2013eva,Li:2016uzn}. It explicitly breaks scale symmetry and generates the scalar-meson mass.

The conformal compensator $\chi$ has canonical mass dimension one and transforms as a field of dimension one under scale transformations, whereas the chiral field $U$ is dimensionless and invariant under scale transformations.
The corresponding scale variation of the Lagrangian gives the divergence of the dilatation current, which describes the scale symmetry breaking as
\begin{equation}
\begin{split}
\partial_\mu j^\mu_D &= 
\alpha \tau\left(\frac{\chi}{f_\sigma}\right)^{4+\alpha}
\left(
{\rm det } U +
{\rm det } U^\dagger -2
\right)
\\
&
- \frac{f_\sigma^2 m_\sigma^2}{4}\left(\frac{\chi}{f_\sigma}\right)^4.
\end{split}
\end{equation}
Furthermore, in connection with the dilatation current, we now introduce the improved EMT,
\begin{equation}
\begin{split}
\Theta^{\mu\nu} &=
\frac{f_\pi^2}{2}
\left(\frac{\chi}{f_\sigma}\right)^2
{\rm tr}\left[
\partial^\mu U \partial^\nu U^\dagger
\right]
+\partial^\mu \chi \partial^\nu\chi
-g^{\mu\nu} {\cal L}
\\
&
-\frac{1}{6}
(\partial^\mu \partial^\nu - g^{\mu\nu} \partial^2 )(\chi^2).
\end{split}
\end{equation}
The last term is the improvement term, as discussed in Subsec.~\ref{EMT_LSM_subsec}.
%which ensures that the trace of the EMT satisfies $\Theta^\mu_\mu = \partial_\mu j^\mu_D$

Using this EMT, we evaluate the pion and $\eta_0$-meson D-form factors within the scalar-meson dominance picture. We obtain
\begin{equation}
\begin{split}
D_{\pi}^{\rm ChPT}(t)&=
-1 +
\frac{2}{3}
\frac{t}{t + m_\sigma^2},\\
D_{\eta_0}^{\rm ChPT}(t)&=D_{\pi}^{\rm ChPT}(t)
-
\frac{2}{3}
\frac{(2+\alpha)m_{\eta_0}^2}{t + m_\sigma^2}.
%\\&=-1 +\frac{2}{3}\frac{t-(2+\alpha)m_{\eta_0}^2}{t + m_\sigma^2}.
\end{split}
\label{eq46}
\end{equation}
The pion D-form factor takes the same form as that obtained in the linear sigma model in the chiral limit:
\begin{equation}
D_{\pi}^{\rm ChPT}(t)
=
\lim_{m_{l,s}=0} D_\pi^{SU(3)_f}(t).
\end{equation}
In contrast, the $\eta_0$-meson D-form factor differs from the linear-sigma-model result. In the chiral limit, the LSM result is given by (see Eq.~(\ref{D_eta0}))
\begin{equation}
\begin{split}
\lim_{m_{l,s}=0}
D_{\eta_0}^{SU(3)_f }(t)
&=
\lim_{m_{l,s}=0} D_\pi^{SU(3)_f }(t)
\\
&-\frac{2}{3}
\frac{
m_{\eta_0}^2 
}{t+ m_{\sigma_0}^2}.
%\label{D_eta0}
\end{split}
\end{equation} 
The difference between the two results originates from the canonical mass dimension of the $U(1)_A$ anomaly interaction.
In the LSM, the determinant operator has canonical mass dimension three, whereas in the present ChPT setup the corresponding $U(1)_A$ anomaly operator is treated as having canonical mass dimension $4+\alpha$ through the factor $(\chi/f_\sigma)^{4+\alpha}$.
When $\alpha=-1$, these two canonical mass dimensions coincide, and the coefficient of the $m_{\eta_0}^2$ term of Eq.~(\ref{eq46}) agrees with the linear-sigma-model result.
Away from this value, the coefficient changes, indicating that the topological contribution to the $\eta_0$-meson D-form factor is sensitive to the canonical mass dimension of the interaction term responsible for the $U(1)_A$ anomaly.

Next, we consider the D-form factor of the lightest scalar meson.
In the present ChPT setup, it is given by
\begin{equation}
\begin{split}
D_\sigma^{\rm ChPT}(t)&= -\frac{11}{3} 
+\frac{1}{3}
\frac{10t}{t + m_\sigma^2}
\end{split}
\end{equation}
On the other hand, the LSM result in the chiral limit is given by
\begin{equation}
\begin{split}
\lim_{m_{l,s}=0}
D_{\sigma_0}^{SU(3)_f }(t)
%&=-1 +\frac{2}{3}\frac{t}{t + m_\sigma^2}
%\\
%&-\frac{2}{9}\frac{ 6m^2_{\sigma_0}+ m^2_{\eta_0}}{t+ m^2_{\sigma_0}}
&=
 -\frac{7}{3}
+
\frac{2t}{t + m^2_{\sigma_0}}
-\frac{2}{9}
\frac{
m^2_{\eta_0}
}{t+ m^2_{\sigma_0}}
%\label{D_SI_t}
\end{split}
\end{equation}

In contrast to the LSM result, the ChPT result does not contain a contribution from the $U(1)_A$ anomaly term.
This difference originates from the model construction of ChPT with a dilatonic scalar particle: in the present setup, the lightest scalar-meson mass is generated solely by the logarithmic dilaton potential $V_{\sigma}^{\rm (ChPT)}$.
However, even if the $U(1)_A$ anomalous contribution is ignored, a crucial difference remains between the LSM and ChPT results.
This difference will be discussed in more detail in the next subsection.

%%%%%%%%%%%%%%%%%%%%%%%%%%%%%%%%%%%%%
\subsection{General result for the scalar-meson D-form factor}

To clarify the origin of the difference in the scalar-meson D-form factor, we consider a simple potential for the lightest scalar-meson field $\phi$, whose vacuum expectation value is given by $\langle 0|\phi|0\rangle=\phi_0$.
The potential is assumed to contain a scale-invariant quartic term and a term with scaling power $\Delta$:
\begin{equation}
\begin{split}
V = C_4 \phi^4 + C_\Delta \phi^\Delta,
\end{split}
\end{equation}
where $C_4$ and $C_\Delta$ are the coupling constants of the quartic term and the $\Delta$-power term, respectively. The second term is necessary to realize a nontrivial vacuum.

The stationary condition at $\phi=\phi_0$ gives
\begin{equation}
\begin{split}
\frac{\partial V}{\partial \phi}
\Biggl|_{\phi=\phi_0}
= 4C_4 \phi_0^3  + \Delta  C_\Delta \phi_0^{\Delta -1} =0.
\end{split}
\end{equation}
Furthermore, the scalar mass and the three point self coupling are obtained as
\begin{equation}
\begin{split}
m_\phi^2&\equiv
\frac{\partial^2 V}{\partial \phi^2}\Biggl|_{\phi=\phi_0}
=\Delta (\Delta -4) C_\Delta\phi_0^{\Delta -2}
, \\
g_{\phi^3} 
&\equiv
\frac{\partial^3 V}{\partial \phi^3}\Biggl|_{\phi=\phi_0} =  (\Delta +1 )m_\phi^2
\phi_0^{-1}
.
\end{split}
\end{equation}
In deriving the final forms of these expressions, we have used the stationary condition. 
Using these quantities under the assumption of lightest-scalar-meson dominance, the corresponding scalar D-form factor can be written as
\begin{equation}
\begin{split}
D_\phi(t) 
&= \frac{-3-2\Delta}{3}
+
\frac{2}{3}
\frac{(\Delta+1) t}{t + m_\phi^2}.
\end{split}
\end{equation}
This expression shows that the scalar D-form factor is sensitive to the scaling power $\Delta$ of the potential.

We now consider several limiting cases.
For $\Delta=2$, the potential reduces to the usual quadratic and quartic form of the linear sigma model.
In this case, one obtains the LSM result for the scalar-meson D-form factor in the absence of the $U(1)_A$ anomaly contribution.

On the other hand, when $\Delta=4+\epsilon$ with $\epsilon\ll1$, 
the potential term proportional to $\phi^\Delta$
generates a logarithmic form, such as $V_{\sigma}^{\rm (ChPT)}$.
In this case, the ChPT result for the scalar-meson D-form factor is reproduced.

Thus, apart from the $U(1)_A$ anomaly contribution, the scalar-meson D-form factor is closely related to the form of the scalar potential, which reflects the canonical mass dimension of the interaction term in the scalar sector.

%%%%%%%%%%%%%%%%%%%%%%%%%%%%%%%%%%%%%
\section{Summary}

In this work, we investigated meson D-form factors in low-energy QCD.
Motivated by the scalar-meson dominance picture, we related the scalar-meson exchange contribution to meson masses through scalar-meson couplings, thereby connecting meson D-form factors to symmetry breaking in low-energy QCD.

To examine this connection, we employed the three-flavor LSM, which incorporates spontaneous chiral symmetry breaking, explicit chiral symmetry breaking, flavor symmetry breaking, and the $U(1)_A$ anomaly.
In the realistic case, flavor symmetry breaking induces singlet-octet mixing, leading to contributions from both $f_0(500)$ and $f_0(980)$ exchange.
%To clarify the role of each symmetry-breaking effect, we first considered the three-flavor symmetric limit, where only the lightest scalar-meson exchange contributes.

In the three-flavor symmetric limit, we derived analytical expressions for the meson D-form factors.
For the octet pseudoscalar mesons, the D-term is fixed to $-1$ in the chiral limit, consistently with the previous study.
On the other hand, the $\eta_0$-meson D-term receives an additional negative contribution from the $U(1)_A$ anomaly, which is related to the topological susceptibility through the Witten--Veneziano relation.
The D-terms of scalar mesons are also affected by the $U(1)_A$ anomaly.
%We also analyzed the scalar-meson D-form factors in the same framework. In the chiral limit, the scalar D-terms are also affected by the $U(1)_A$ anomaly through the $\eta_0$-meson mass. For the scalar-octet mesons, the D-terms also depend on the scalar-octet mass, represented by $m_\delta$.

We then numerically evaluated the meson D-form factors for the realistic case $m_l\neq m_s$ using two parameter sets.
One set excludes the determinant-type interaction involving the current quark mass, $V_{\rm SB-anom}$, whereas the other includes it to reproduce the scalar-meson mass hierarchy below $1\,{\rm GeV}$.
The pseudoscalar D-form factors show similar qualitative behavior for the two sets.
In contrast, the scalar-meson D-form factors exhibit a strong dependence on the parameter $k$ associated with $V_{\rm SB-anom}$.
In particular, the ordering of the D-term of $f_0(500)$ changes significantly when a nonzero value of $k$ is introduced.
%The detailed ordering of the other scalar-meson D-terms is also affected by the singlet-octet scalar-meson mixing.
%These results indicate that the scalar-meson D-form factors are especially sensitive to the $V_{\rm SB-anom}$ and scalar-meson mixing.

Finally, we compared the LSM results with those from ChPT including a dilatonic scalar field. While the pion D-form factor agrees in the chiral limit, the $\eta_0$
and lightest-scalar D-form factors differ because the relevant interaction terms carry different canonical mass dimensions in the two approaches. This shows that meson D-form factors are sensitive not only to meson masses but also to the scaling properties of the interactions responsible for mass generation.

%Finally, we compared the LSM results with those obtained from ChPT including a dilatonic scalar field. For the pion D-form factor, the ChPT result agrees with the LSM result in the chiral limit.For the $\eta_0$ meson, however, the coefficient of the $U(1)_A$ anomaly contribution depends on the way the $U(1)_A$ anomalous interaction is implemented. In the LSM, the determinant interaction has canonical mass dimension three, whereas in the present ChPT setup the corresponding anomalous term is characterized by the power $4+\alpha$ through the conformal compensator. This difference changes the coefficient of the $U(1)_A$ anomaly contribution to the $\eta_0$-meson D-form factor. Thus, the topological contribution to the $\eta_0$-meson D-form factor is sensitive to the canonical mass dimension associated with the $U(1)_A$ anomalous interaction. Similarly, the comparison for the D-form factor of the lightest scalar meson shows that, although the $U(1)_A$ anomaly contribution is present, the difference between the ChPT and LSM results is also related to the different canonical mass dimensions of the interaction terms in the scalar-meson potential. Taken together, these results imply that meson D-form factors are closely connected not only to the meson masses themselves but also to the canonical mass dimensions of the interaction terms responsible for meson mass generation.

In this study, we focused on the low-energy region of QCD below about $1~{\rm GeV}$, where the lightest scalar-meson dominance picture is expected to be relevant.
At higher momentum transfer, other hadronic exchanges, such as tensor-meson exchange, may also become important~\cite{Broniowski:2024oyk}.
Such contributions may affect not only the D-form factor but also other gravitational form factors, including the A-form factor.
We leave the study of these higher-energy contributions for future work.

Our results suggest that meson D-form factors provide sensitive
probes of chiral, flavor, $U(1)_A$, and scale-symmetry breaking
in low-energy hadron structure and may serve as useful benchmarks
for future lattice QCD and phenomenological studies.

%%%%%%%%%%%%%%%%%%%%%%%%%%%%%%%%%%%%%

\acknowledgments

The work of M.K. is supported by RFIS-NSFC under Grant No. W2433019.
The work of K.T. is supported by JSPS KAKENHI Grant Numbers JP24K07055 and JP23K03419.
The work of M.T. is supported by ``THERS Make New Standards Program for the Next Generation Researchers" and JST SPRING, Grant Number JPMJSP2125.

\bibliography{reference}

\end{document}